\documentclass[twocolumn,prl,showpacs,superscriptaddress]{revtex4}

\usepackage{graphicx}
\usepackage{dcolumn}
\usepackage{bm}

\begin{document}

\title{The Composite Particle-Hole Spinor of the Lowest Landau Level}
\author{Jian Yang}
\email{jyangmay1@yahoo.com}
\altaffiliation{Permanent address: 5431 Chesapeake Place, Sugar Land, TX 77479, USA}
%\affiliation{}
%\date{}

\begin{abstract}
We propose to form a two-component effective field theory from $\mathcal{L} = \frac{1}{2}(\mathcal{L}_{ce} + \mathcal{L}_{ch})$, where $\mathcal{L}_{ce}$ is the Lagrangian of composite electrons with a Chern-Simons term, and $\mathcal{L}_{ch}$ is the particle-hole conjugate  of $\mathcal{L}_{ce}$ - the Lagrangian of composite holes. It is shown $\mathcal{L} = {\psi}^{\dagger}(i{\partial}_{0}-a_{0}){\psi}-\frac{1}{2m^*}{\psi}^{\dagger}[\sigma_{i}(i{\partial}_{i}-a_{i})]^2{\psi} +\frac{1}{4\pi}{\epsilon}^{\mu \nu \lambda}A_{\mu}{\partial}_{\nu}a_{\lambda}+\frac{1}{8\pi}{\epsilon}^{\mu \nu \lambda}A_{\mu}{\partial}_{\nu}A_{\lambda}$, where the two-component fermion field ${\psi}$ is a composite particle-hole spinor coupled to an emergent effective gauge field $a_{\mu}$ in the presence of a background electromagnetic field $A_{\mu}$. In the theory, the Chern-Simons terms for both the composite electrons and composite holes are exactly cancelled out, and a $\frac{1}{2}$ pseudospin degree of freedom, which responses to the emergent gauge field the same way as the real spin to the electromagnetic field, emerges automatically. Furthermore, the composite particle-hole spinor theory has exactly the same form as the non-relativistic limit of the massless Dirac composite fermion theory after expanded to the four-component form and with a mass term added.
\end{abstract}

 \maketitle

\section{\label{sec:level1}I. INTRODUCTION}

One of the most remarkable proposals in the study of the fractional quantum Hall effect was formulated by Halperin, Lee, and Read some twenty year ago\cite{HLR}, where the two-dimensional electrons in a strong magnetic field at exact half filling of the lowest Landau level behave like a compressible Fermi liquid at zero magnetic field. According to the theory which will be referred to as the HLR theory in this paper, the composite electrons are formed by attaching two flux quanta to each of the original electrons. In the mean field approximation, the attached flux cancels the external magnetic field exactly at half filling, resulting in a Fermi liquid description of the composite fermions. It is generally believed that the HLR theory lacks particle-hole (PH) symmetry\cite{Kivelson}, even though the two-body interaction Hamiltonian of the original electrons when projected onto the lowest Landau level is invariant by an antiunitary PH transformation at half-filled Landau level. This means the PH symmetry is spontaneously broken,  the HLR theory describes only the low-energy excitations around one of the two ground states and a distinct PH conjugate  state of the composite electron state exists. This leads to the proposal of a PH conjugate theory by Barkeshli, Mulligan, and Fisher\cite{BMF}, which is referred as BMF theory in this paper, by forming composite holes with two units of flux quanta attached to each of the holes of the filled Landau level instead of the electrons. It is argued that the BMF theory describes a distinct state of matter as compared to the HLR theory\cite{BMF}. 

On the other hand, the finite size numerical results seem to confirm the PH symmetry of the ground state at the half-filled Landau level\cite{Rezayi}\cite{Geraedts}, and a PH symmetric Dirac composite fermion theory is proposed by Son\cite{Son}\cite{Son1}. In this theory, a massless Dirac composite fermion is characterized by a Berry phase of $\pi$ around the Fermi circle, and is interpreted as a type of fermionic vortex, arising from a fermionic particle-vortex duality. A key feature of particle-vortex duality is that it switches the roles of particle number and magnetic field, and therefore the Dirac composite fermion density is set by the external magnetic field, which is different from the density of the original electrons when away from the half-filling. Very recently, a similar but different Dirac composite fermion effective field theory is proposed by the author\cite{DiracSpinor}, where the two-component Dirac composite fermion field is a particle-hole spinor coupled to the same emergent gauge field, with one field component describing the composite electrons and the other describing the PH conjugated composite holes. As such, the density of the Dirac spinor field is the density sum of the composite electron and hole field components, and therefore is equal to the degeneracy of the Lowest Landau level. On the other hand, the charge density coupled to the external magnetic field is the density difference between the composite electron and hole field components, and is therefore neutral at exactly half-filling. The two Dirac composite theories are argued to give essentially the same electromagnetic response, although their exact relationship remains to be determined. 

Furthermore, the Dirac composite fermion theories have not been shown to be derived microscopically from the original electron Hamiltonian. In contrast, both HLR theory and BMF theory are considered to be a mathematical transformation from the original microscopic Hamiltonian with flux attachment encoded in the Chern-Simons term. 

We start from the HLR's Chern-Simons Lagrangian of composite electrons formed by attaching two flux quanta to each of the electrons \cite{HLR}
\begin{equation}
\label{CELagrangian}
\begin{array}{l@{}l}
\mathcal{L}_{ce} = {\psi}^{\dagger}_{ce}(i{\partial}_{0}+A_{0}+c_{0} + {\mu}_{ce}){\psi}_{ce}\\
-\frac{1}{2m^*}{\psi}^{\dagger}_{ce}(i{\partial}_{i}+A_{i}+c_{i})^2{\psi}_{ce} +\frac{1}{8\pi}{\epsilon}^{\mu \nu \lambda}c_{\mu}{\partial}_{\nu}c_{\lambda}
\end{array}
\end{equation}
where ${\psi}_{ce}$ is the composite electron field, $A_{\mu}$ is the external electromagnetic field, $c_{\mu}$ is the Chern-Simons gauge field with the Chern-Simons term $\frac{1}{8\pi}{\epsilon}^{\mu \nu \lambda}c_{\mu}{\partial}_{\nu}c_{\lambda}$, ${\mu}_{ce}$ is the chemical potential, and $m^*$ is the effective mass of the composite electron.

On the other hand, the BMF's Chern-Simons Lagrangian of composite holes, formed by attaching two flux quanta to each of the holes of the filled Landau level of the electrons, is \cite{BMF}
\begin{equation}
\label{CHLagrangian}
\begin{array}{l@{}l}
\mathcal{L}_{ch} = {\psi}^{\dagger}_{ch}(i{\partial}_{0}-A_{0}+d_{0} + {\mu}_{ch}){\psi}_{ch}\\
-\frac{1}{2m^*}{\psi}^{\dagger}_{ch}(i{\partial}_{i} -A_{i}+d_{i})^2{\psi}_{ch} 
-\frac{1}{8\pi}{\epsilon}^{\mu \nu \lambda}d_{\mu}{\partial}_{\nu}d_{\lambda} \\
+ \frac{1}{4\pi}{\epsilon}^{\mu \nu \lambda}A_{\mu}{\partial}_{\nu}A_{\lambda}
\end{array}
\end{equation}
where ${\psi}_{ch}$ is the composite hole field, $A_{\mu}$ is the same external electromagnetic field, $d_{\mu}$ is the Chern-Simons gauge field with the Chern-Simons term $-\frac{1}{8\pi}{\epsilon}^{\mu \nu \lambda}d_{\mu}{\partial}_{\nu}d_{\lambda}$, ${\mu}_{ch}$ is the chemical potential, and $m^*$ is the effective mass of the composite hole which is assumed to be the same as the composite electron.

Comparing Eq. (\ref{CHLagrangian}) with Eq. (\ref{CELagrangian}), we notice a few differences. The first difference is the opposite signs in the external electromagnetic field $A$ term coupled to the matter field, which reflects the opposite charges of the composite electron and the composite hole. The second difference is the opposite sign of the Chern-Simons term, which is required as the attached flux quanta are such that they are opposite to the effect of the external magnetic field seen by the composite electrons and composite holes. The third difference is the Chern-Simons term of the background electromagnetic field $\frac{1}{4\pi}{\epsilon}^{\mu \nu \lambda}A_{\mu}{\partial}_{\nu}A_{\lambda}$ appeared in Eq. (\ref{CHLagrangian}). This term is necessary to ensure the electromagnetic response of $\mathcal{L}_{ch}$ in Eq. (\ref{CHLagrangian}) for the vacuum state of the composite holes gives a correct result of a filled Landau level of the electrons.

\section{\label{sec:level1}II. COMPOSITE PARTICLE-HOLE SPINOR EFFECTIVE FIELD THEORY}

We now make a variable replacement
\begin{equation}
c_{\mu} + A_{\mu} = -{c'}_{\mu}
\end{equation}
to rewrite the Lagrangian $\mathcal{L}_{ce}$ in an equivalent form
\begin{equation}
\label{CELagrangian1}
\begin{array}{l}
\mathcal{L}_{ce} ={\psi}^{\dagger}_{ce}(i{\partial}_{0}-c'_{0}+{\mu}_{ce}){\psi}_{ce}-\frac{1}{2m^*}{\psi}^{\dagger}_{ce}(i{\partial}_{i}-c'_{i})^2{\psi}_{ce} \\
+\frac{1}{8\pi}{\epsilon}^{\mu \nu \lambda}c'_{\mu}{\partial}_{\nu}c'_{\lambda}+\frac{1}{8\pi}{\epsilon}^{\mu \nu \lambda}A_{\mu}{\partial}_{\nu}A_{\lambda} + \frac{1}{4\pi}{\epsilon}^{\mu \nu \lambda}A_{\mu}{\partial}_{\nu}c'_{\lambda}
\end{array}
\end{equation}
Similarly, by making a variable replacement
\begin{equation}
d_{\mu} - A_{\mu} = -{d'}_{\mu}
\end{equation}
the Lagrangian $\mathcal{L}_{ch}$ becomes
\begin{equation}
\label{CHLagrangian1}
\begin{array}{l}
\mathcal{L}_{ch} = {\psi}^{\dagger}_{ch}(i{\partial}_{0}-d'_{0}+{\mu}_{ch}){\psi}_{ch}-\frac{1}{2m^*}{\psi}^{\dagger}_{ch}(i{\partial}_{i}-d'_{i})^2{\psi}_{ch} \\
-\frac{1}{8\pi}{\epsilon}^{\mu \nu \lambda}d'_{\mu}{\partial}_{\nu}d'_{\lambda}+\frac{1}{8\pi}{\epsilon}^{\mu \nu \lambda}A_{\mu}{\partial}_{\nu}A_{\lambda} + \frac{1}{4\pi}{\epsilon}^{\mu \nu \lambda}A_{\mu}{\partial}_{\nu}d'_{\lambda}
\end{array}
\end{equation}

From the equation of motion for $c'_0$ using $\mathcal{L}_{ce}$, we have
\begin{equation}
\label{CEDensity}
\psi_{ce}^{\dagger} \psi_{ce} = \frac{B}{4\pi} - \frac{b_{ce}}{4\pi}
\end{equation}
where $B = {\epsilon}_{ij}{\partial}_{i}A_{j}$, $b_{ce} = {\epsilon}_{ij}{\partial}_{i}c'_{j}$, and ${\epsilon}_{ij}$ is the anti-symmetric unit tensor. Similarly from the equation of motion for $d'_0$ using $\mathcal{L}_{ch}$, we have
\begin{equation}
\label{CHDensity}
\psi_{ch}^{\dagger} \psi_{ch} = \frac{B}{4\pi} + \frac{b_{ch}}{4\pi}
\end{equation}
where $b_{ch} = {\epsilon}_{ij}{\partial}_{i}d'_{j}$. Since $\psi_{ce}$ and $\psi_{ch}$ are the PH conjugate fields, they must satisfy the following equation
\begin{equation}
\label{DensitySum}
\psi_{ch}^{\dagger} \psi_{ch} + \psi_{ce}^{\dagger} \psi_{ce} = \frac{B}{2\pi}
\end{equation}
This requires 
\begin{equation}
\label{PHCondition}
b_{ch} = b_{ce} = b
\end{equation}
or more generally
\begin{equation}
\label{PHCondition1}
{c'}_{\mu} = {d'}_{\mu} = {a}_{\mu}
\end{equation}
where $b = {\epsilon}_{ij}{\partial}_{i}a_{j}$. 

From Eq. (\ref{CEDensity}) and Eq. (\ref{CHDensity}), using Eq. (\ref{PHCondition}), we have 
\begin{equation}
\label{DensityDifference}
\psi_{ch}^{\dagger} \psi_{ch} - \psi_{ce}^{\dagger} \psi_{ce} = \frac{b}{2\pi}
\end{equation}
This means that the composite hole density is always larger (or smaller) than the composite electron density by the amount of $\frac{|b|}{2\pi}$ when $b>0$ ($b<0$). Since $\frac{|b|}{2\pi}$ is the Landau level degeneracy of $b$ magnetic field, this means that the composite holes always occupy one more (one less) Landau level than the composite electrons when $b>0$ ($b<0$). This is only possible when the chemical potential of the composite electrons $\mu_{ce}$ is larger (smaller) than the chemical potential of the composite holes by exactly one Landau level gap $\frac{|b|}{m^*}$ when $b>0$ ($b<0$). Without a loss of generality, this means we can write
\begin{equation}
\label{ChemicalPotential}
\begin{array}{l}
\mu_{ce} = \mu+\mu_b\\
\mu_{ch} = \mu-\mu_b\\
\mu_b = \frac{b}{2m^*}
\end{array}
\end{equation}
where the chemical potential $\mu_b$ is in general position dependent as the emergent gauge field is in general position dependent.

Since $\mathcal{L}_{ce}$ and $\mathcal{L}_{ch}$ given by Eq. (\ref{CELagrangian1}) and Eq. (\ref{CHLagrangian1}) are PH conjugate with each other under condition Eq. (\ref{PHCondition1}) and Eq. (\ref{ChemicalPotential}), we propose the following PH symmetric effective field theory
\begin{equation}
\mathcal{L} = \frac{1}{2}(\mathcal{L}_{ce} + \mathcal{L}_{ch})
\end{equation}
which is 
\begin{equation}
\label{PauliLagrangian}
\begin{array}{l@{}l}
\mathcal{L} = {\psi}^{\dagger}(i{\partial}_{0}-a_{0}){\psi}-\frac{1}{2m^*}{\psi}^{\dagger}(i{\partial}_{i}-a_{i})^2{\psi} + \frac{b}{2m^*}{\psi}^{\dagger}{\sigma}_3{\psi}\\
+\frac{1}{4\pi}{\epsilon}^{\mu \nu \lambda}A_{\mu}{\partial}_{\nu}a_{\lambda}+\frac{1}{8\pi}{\epsilon}^{\mu \nu \lambda}A_{\mu}{\partial}_{\nu}A_{\lambda}
\end{array}
\end{equation}
where 
\begin{equation}
\psi = \frac{1}{\sqrt 2}\left(
\begin{array}{c}
\psi_{ch}\\
\psi_{ce}
\end{array}\right)\
\end{equation}
$\sigma_{i}$ are the Pauli matrices. Notice the Chern-Simon terms in Eq. (\ref{CELagrangian1}) and Eq. (\ref{CHLagrangian1}) are cancelled out exactly, and the Stern--Gerlach term $\frac{b}{2m^*}{\psi}^{\dagger}{\sigma}_3{\psi}$ is appeared. For simplicity, we have dropped out the common chemical potential $\mu$ in Eq. (\ref{PauliLagrangian}). One can rewrite Eq. (\ref{PauliLagrangian}) into a more compact form 
\begin{equation}
\label{PauliLagrangian1}
\begin{array}{l@{}l}
\mathcal{L} = {\psi}^{\dagger}(i{\partial}_{0}-a_{0}){\psi}-\frac{1}{2m^*}{\psi}^{\dagger}[\sigma_{i}(i{\partial}_{i}-a_{i})]^2{\psi} \\
+\frac{1}{4\pi}{\epsilon}^{\mu \nu \lambda}A_{\mu}{\partial}_{\nu}a_{\lambda}+\frac{1}{8\pi}{\epsilon}^{\mu \nu \lambda}A_{\mu}{\partial}_{\nu}A_{\lambda}
\end{array}
\end{equation}
As is clearly seen, a $\frac{1}{2}$ pseudospin degree of freedom, which response to the emergent gauge field the same way as the real spin to the electromagnetic field, emerges automatically from the theory. 

The equation of motion for $a_0$ will give the density operator of the particle-hole spinor
\begin{equation}
\label{EquationOfMotion_a0}
\rho = \psi^{\dagger} \psi  = \frac{B}{4\pi}
\end{equation}
On the other hand, by differentiating  the action with respect to $A_0$, we obtain the electron density
\begin{equation}
\label{ElectronDensity}
\rho^e = \frac{B}{4\pi}-\frac{b}{4\pi}
\end{equation}
Similar to the charge density equations Eq. (\ref{EquationOfMotion_a0}) and Eq. (\ref{ElectronDensity}), one can obtain the current density equations from the equation of motion for $a_i$  
\begin{equation}
\label{EquationOfMotion_a}
j_i = \frac{1}{4\pi}{\epsilon}_{ij}E_j
\end{equation}
By differentiating the action with respect to $A_i$, we obtain
\begin{equation}
\label{ElectronCurrent}
j^e_i = \frac{1}{4\pi}{\epsilon}_{ij}(E_j-e_j)
\end{equation}
where  $j$ and $j^e$ represent  the current densities  of composite particle-hole spinor field and the electron field respectively, $e_j = -{\partial}_{ij}a_0$ and $E_j = -{\partial}_{ij}A_0$ are the emergent and external electric fields respectively.  

It is noticed, by replacing $(\rho,j)$ with the corresponding relativistic density-current vector, Eq. (\ref{EquationOfMotion_a0}), Eq. (\ref{ElectronDensity}), Eq. (\ref{EquationOfMotion_a}), and Eq. (\ref{ElectronCurrent}) remain valid from the following Dirac composite fermion theory proposed by Son\cite{Son}\cite{Son1} 
\begin{equation}
\label{SonLagrangian}
\mathcal{L}_d = i\bar{\psi}_d{\gamma}^{\mu}({\partial}_{\mu}+ia_{\mu})\psi_d+\frac{1}{4\pi}{\epsilon}^{\mu \nu \lambda}A_{\mu}{\partial}_{\nu}a_{\lambda}+\frac{1}{8\pi}{\epsilon}^{\mu \nu \lambda}A_{\mu}{\partial}_{\nu}A_{\lambda}
\end{equation}
In fact, if we make $\mathcal{L}$ in Eq. (\ref{PauliLagrangian1}) linear in ${\partial}_{i}$ by dropping out $\frac{1}{2m^*}\sigma_{i}(i{\partial}_{i}-a_{i})$ from the second term, the Lagrangian $\mathcal{L}$ becomes
\begin{equation}
\label{PauliLagrangian2}
\begin{array}{l@{}l}
\mathcal{L} \rightarrow {\psi}^{\dagger}(i{\partial}_{0}-a_{0}){\psi}-{\psi}^{\dagger}\sigma_{i}(i{\partial}_{i}-a_{i}){\psi} \\
+\frac{1}{4\pi}{\epsilon}^{\mu \nu \lambda}A_{\mu}{\partial}_{\nu}a_{\lambda}+\frac{1}{8\pi}{\epsilon}^{\mu \nu \lambda}A_{\mu}{\partial}_{\nu}A_{\lambda}
\end{array}
\end{equation}
which is identical to $\mathcal{L}_d$ in Eq. (\ref{SonLagrangian}). On the other hand, if we expand the massless Dirac composite fermion theory $\mathcal{L}_d$ to a four component form and add a mass term to it, one can show its non-relativistic limit takes exactly the same form as the composite particle-hole spinor theory $\mathcal{L}$ in Eq. (\ref{PauliLagrangian1}).  

\section{\label{sec:level1}III. ALTERNATIVE COMPOSITE PARTICLE-HOLE SPINOR EFFECTIVE FIELD THEORY}

As was pointed out previously, compared to Eq. (\ref{CELagrangian}), there is a background electromagnetic term  $\frac{1}{4\pi}{\epsilon}^{\mu \nu \lambda}A_{\mu}{\partial}_{\nu}A_{\lambda}$ in Eq. (\ref{CHLagrangian}). This term is necessary to ensure the electromagnetic response of $\mathcal{L}_{ch}$ in Eq. (\ref{CHLagrangian}) for the vacuum state of the composite holes gives a correct result of a filled Landau level of the electrons. Without this background term, the Lagrangian $\mathcal{L}_{ch}$ would describe the composite holes whose electromagnetic response will differ by that of a  filled Landau level of electrons. If we remove this term from Eq. (\ref{CHLagrangian}), and add it to Eq. (\ref{CELagrangian}) to form the following Lagrangian 
\begin{equation}
\mathcal{L'} = \mathcal{L}_{ce} + \mathcal{L}_{ch} - \frac{1}{4\pi}{\epsilon}^{\mu \nu \lambda}A_{\mu}{\partial}_{\nu}A_{\lambda}
\end{equation}
which is related to $\mathcal{L}$ in Eq. (\ref{PauliLagrangian1}) by
\begin{equation}
\mathcal{L'} = 2 \mathcal{L} - \frac{1}{4\pi}{\epsilon}^{\mu \nu \lambda}A_{\mu}{\partial}_{\nu}A_{\lambda}
\end{equation}
or 
\begin{equation}
\label{PauliLagrangian3}
\begin{array}{l@{}l}
\mathcal{L'} = {\phi}^{\dagger}(i{\partial}_{0}-a_{0}){\phi}-\frac{1}{2m^*}{\phi}^{\dagger}[\sigma_{i}(i{\partial}_{i}-a_{i})]^2{\phi} \\
+\frac{1}{2\pi}{\epsilon}^{\mu \nu \lambda}A_{\mu}{\partial}_{\nu}a_{\lambda}
\end{array}
\end{equation}
where 
\begin{equation}
\phi = \left(
\begin{array}{c}
\psi_{ch}\\
\psi_{ce}
\end{array}\right)\
\end{equation}

The equation of motion for $a_0$ gives the field density operator
\begin{equation}
\label{EquationOfMotion_a01}
\rho' = \phi^{\dagger}\phi = \psi_{ch}^{\dagger} \psi_{ch} + \psi_{ce}^{\dagger} \psi_{ce} = \frac{B}{2\pi}
\end{equation}
On the other hand, by differentiating  the action with respect to $A_0$, and equating the result to $\psi_{ch}^{\dagger} \psi_{ch} - \psi_{ce}^{\dagger} \psi_{ce}$ since $\psi_{ch}$ and $\psi_{ce}$ have opposite electromagnetic charges, we can obtain 
\begin{equation}
\label{EquationOfMotion_A01}
\psi_{ch}^{\dagger} \psi_{ch} - \psi_{ce}^{\dagger} \psi_{ce} = \frac{b}{2\pi}
\end{equation}
which is exactly the same as Eq. (\ref{DensityDifference}). From Eq. (\ref{EquationOfMotion_a01}) and  Eq. (\ref{EquationOfMotion_A01}), we can obtain the electron density operator
\begin{equation}
\label{ElectronDensity1}
{\rho}^e = \psi_{ce}^{\dagger} \psi_{ce} = \frac{B}{4\pi}-\frac{b}{4\pi}
\end{equation}
which is the same as Eq. (\ref{ElectronDensity}).

One can obtain the current density equation $j'$ from the equation of motion for $a_i$  
\begin{equation}
\label{EquationOfMotion_a1}
j'_i = j^{ce}_i + j^{ch}_i = \frac{1}{2\pi}{\epsilon}_{ij}E_j
\end{equation}
By differentiating the action with respect to $A_i$, and equating the result to $j^{ch}_i- j^{ce}_i$ since $\psi_{ch}$ and $\psi_{ce}$ have opposite electromagnetic charges, we can obtain 
\begin{equation}
\label{EquationOfMotion_A1}
 j^{ch}_i- j^{ce}_i = \frac{1}{2\pi}{\epsilon}_{ij}e_j
\end{equation}
Combining Eq. (\ref{EquationOfMotion_a1}) and Eq. (\ref{EquationOfMotion_A1}), we obtain the electron current operator
\begin{equation}
\label{ElectronCurrent1}
j^{e}_i = \frac{1}{4\pi}{\epsilon}_{ij}(E_j-e_j)
\end{equation}
which is exactly the same as Eq. (\ref{ElectronCurrent}). In fact, the set of equations Eq. (\ref{EquationOfMotion_a01}), Eq. (\ref{ElectronDensity1}), Eq. (\ref{EquationOfMotion_a1}), and Eq. (\ref{ElectronCurrent1}) is identical to the set of equations of Eq. (\ref{EquationOfMotion_a0}), Eq. (\ref{ElectronDensity}), Eq. (\ref{EquationOfMotion_a}), and Eq. (\ref{ElectronCurrent}), if we let $\rho' = 2\rho$ and $j' = 2j$. Therefore, we conclude $\mathcal{L'}$ in Eq. (\ref{PauliLagrangian3}) is equivalent to $\mathcal{L}$ in Eq. (\ref{PauliLagrangian1}).

As before, one can obtain the relativistic version of $\mathcal{L'}$ by dropping out $\frac{1}{2m^*}\sigma_{i}(i{\partial}_{i}-a_{i})$ from the second term in Eq. (\ref{PauliLagrangian3})
\begin{equation}
\label{PauliLagrangian4}
\begin{array}{l@{}l}
\mathcal{L'} \rightarrow {\phi}^{\dagger}(i{\partial}_{0}-a_{0}){\phi}-{\phi}^{\dagger}\sigma_{i}(i{\partial}_{i}-a_{i}){\phi} \\
+\frac{1}{2\pi}{\epsilon}^{\mu \nu \lambda}A_{\mu}{\partial}_{\nu}a_{\lambda}
\end{array}
\end{equation}
which is identical to the following Lagrangian which is written in a covariant  form
\begin{equation}
\label{DiracLagrangian}
\mathcal{L'}_d = i\bar{\phi}{\gamma}^{\mu}({\partial}_{\mu}+ia_{\mu})\phi+\frac{1}{2\pi}{\epsilon}^{\mu \nu \lambda}A_{\mu}{\partial}_{\nu}a_{\lambda}
\end{equation}
which was first proposed in reference \cite{DiracSpinor}. Since $\mathcal{L'}$ in Eq. (\ref{PauliLagrangian3}) is equivalent to $\mathcal{L}$ in Eq. (\ref{PauliLagrangian1}), we conclude their relativistic versions in Eq. (\ref{SonLagrangian}) and Eq. (\ref{PauliLagrangian4}) are also equivalent.

\section{\label{sec:level1}IV. CONCLUSION}
We proposed a two-component non-relativistic effective field theory from $\mathcal{L} = \frac{1}{2}(\mathcal{L}_{ce} + \mathcal{L}_{ch})$, where $\mathcal{L}_{ce}$ is the Lagrangian of composite electrons with a Chern-Simons term, and $\mathcal{L}_{ch}$ is the particle-hole conjugate  of $\mathcal{L}_{ce}$ - the Lagrangian of composite holes. We also proposed an alternative non-relativistic effective field theory. Since both non-relativistic effective field theories are equivalent, we conclude their relativistic versions proposed by Son\cite{Son}\cite{Son1} and by the author\cite{DiracSpinor} are also equivalent.


\begin{thebibliography}{10}

\bibitem{HLR} B. I. Halperin, P. A. Lee, and N. Read, Phys. Rev. {\bf B 47},7312 (1993).
\bibitem{Kivelson} S. A. Kivelson, D.-H. Lee, Y. Krotov, and J. Gan, Phys. Rev. {\bf B 55},15552 (1997).
\bibitem{BMF} M. Barkeshli, M. Mulligan, M. P. A. Fisher, Phys. Rev. {\bf B 92}, 165125 (2015).

\bibitem{Rezayi} E. H. Rezayi and F. D. M. Haldane, Phys. Rev. Lett. {\bf84}, 4685 (2000).
\bibitem{Geraedts} S. D. Geraedts, M. P. Zaletel, R. S. K. Mong, M. A. Metlitski, A. Vishwanath, and O. I. Motrunich, Science 352, 197 (2016).

\bibitem{Son} D. T. Son, Phys. Rev. X {\bf5}, 031027 (2015).
\bibitem{Son1} D. T. Son, Prog. of Theor. and Exp. Phys. 12C103 (2016).

\bibitem{DiracSpinor} J. Yang, arXiv:1711.08520

\end{thebibliography}
\end{document}